\documentclass[apj,twocolumn]{emulateapj}
\usepackage{amsmath}
\usepackage{amssymb}

\begin{document}
\shorttitle{\sc The Sgr Streams in the South}

\title{The Sagittarius Streams in the Southern Galactic Hemisphere}
\author{Sergey E. Koposov\altaffilmark{1,2},
V. Belokurov\altaffilmark{1},
N.W. Evans\altaffilmark{1},
G. Gilmore\altaffilmark{1}, 
M. Gieles\altaffilmark{1},
M.J. Irwin\altaffilmark{1},
G.F. Lewis\altaffilmark{1,3},
M. Niederste-Ostholt\altaffilmark{1},
J. Pe{\~n}arrubia\altaffilmark{1},
M.C. Smith\altaffilmark{4},
D. Bizyaev\altaffilmark{5,2},
E. Malanushenko\altaffilmark{5},
V. Malanushenko\altaffilmark{5},
D.P. Schneider\altaffilmark{6},
R.F.G. Wyse\altaffilmark{7}
} 

\altaffiltext{1}{Institute of Astronomy, Madingley Road,
 Cambridge CB3 0HA, UK}
\altaffiltext{2}{Sternberg Astronomical Institute, Moscow State University,
Universitetskiy pr. 13, Moscow 119991, Russia}
\altaffiltext{3}{Sydney Institute for Astronomy, School of Physics,
  A28, The University of Sydney, NSW 2006, Australia}
\altaffiltext{4}{Kavli Institute for Astronomy and Astrophysics,
  Peking University, Beijing 100871, China; National Astronomical
  Observatoires, Chinese Academy of Sciences, Beijing 100012, China}
\altaffiltext{5}{Apache Point Observatory, Sunspot, NM, 88349, USA}
\altaffiltext{6}{Department of Astronomy and Astrophysics, The Pennsylvania
State University, 525 Davey Laboratory, University Park, PA 16802, USA}
\altaffiltext{7}{Department of Physics \& Astronomy, The Johns Hopkins
University, 3900 N. Charles Street, Baltimore, MD 21218, USA}

\begin{abstract}
  The structure of the Sagittarius stream in the Southern Galactic
  hemisphere is analysed with the Sloan Digital Sky Survey Data
  Release 8.  Parallel to the Sagittarius tidal track, but $\sim
  10^\circ$ away, there is another fainter and more metal-poor stream. We
  provide evidence that the two streams follow similar distance
  gradients but have distinct morphological properties and stellar
  populations.  The brighter stream is broader, contains more
  metal-rich stars and has a richer colour-magnitude diagram with
  multiple turn-offs and a prominent red clump as compared to the
  fainter stream. Based on the structural properties and the stellar
  population mix, the stream configuration is similar to the Northern
  ``bifurcation''.  In the region of the South Galactic Cap, there is
  overlapping tidal debris from the Cetus Stream, which crosses the
  Sagittarius stream. Using both photometric and spectroscopic data, we show
that the
  blue straggler population belongs mainly to Sagittarius and the blue
  horizontal branch stars belong mainly to the Cetus stream in this confused
  location in the halo.
\end{abstract}


\maketitle

\begin{figure*}
 \includegraphics{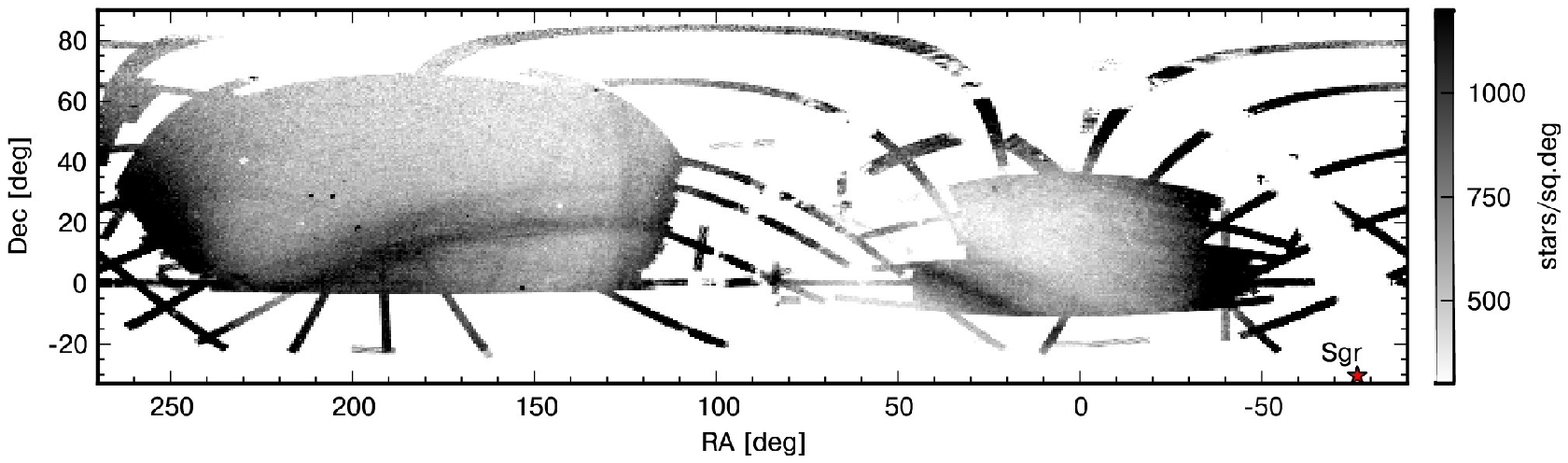}\\
 \includegraphics{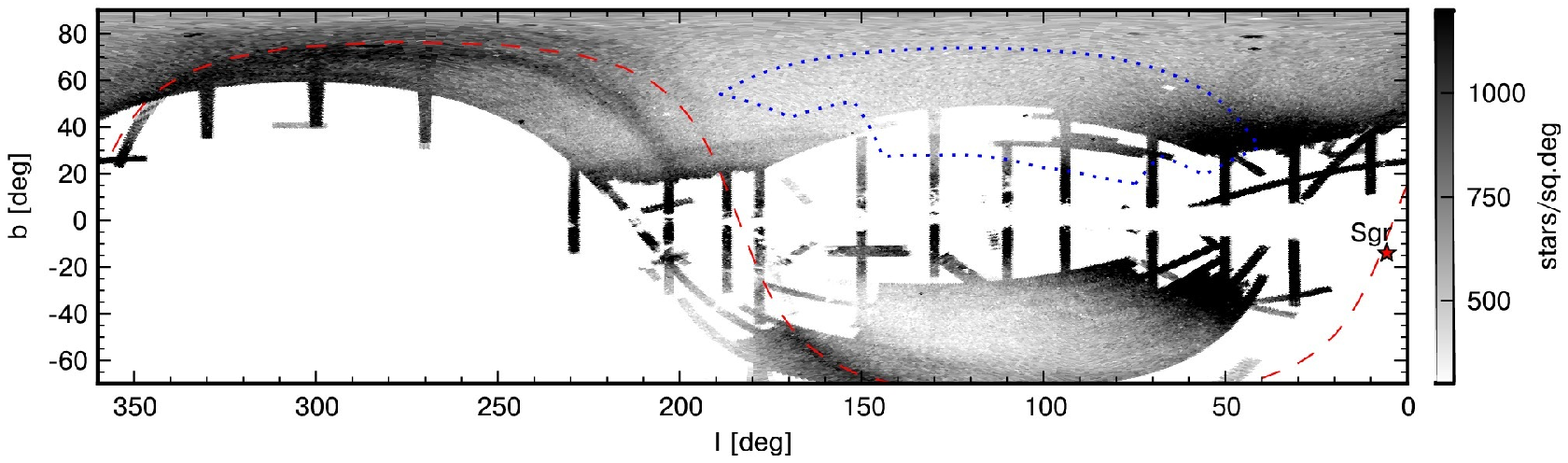}\\
 \includegraphics{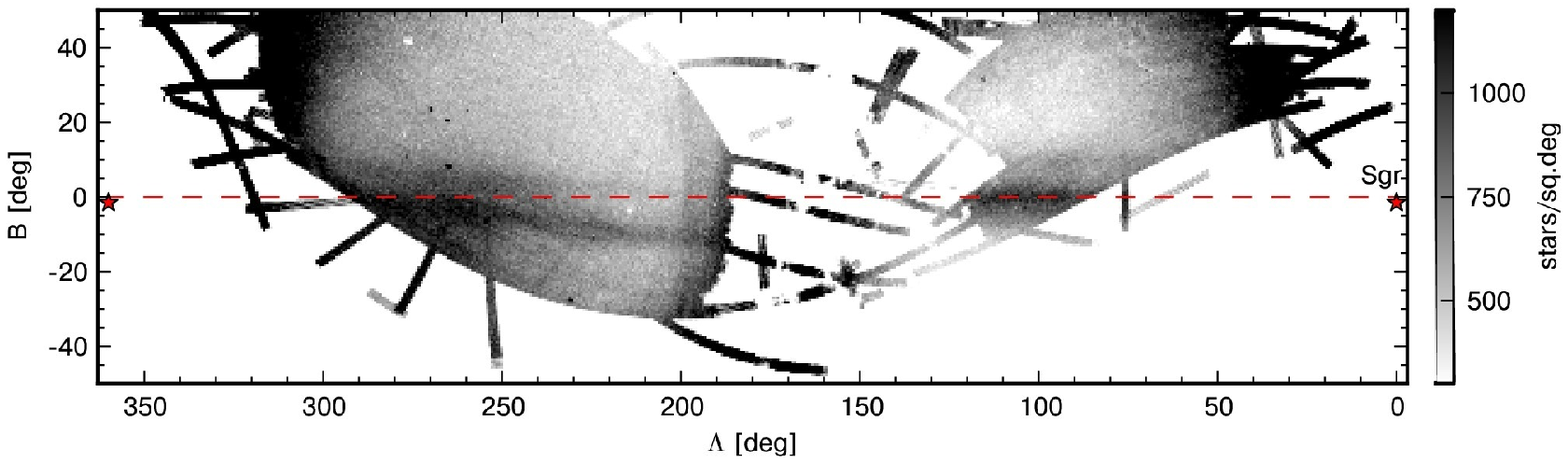}
 \caption{ Density of MSTO stars with $0<g\!-\!i<0.7$ and $19.5<i<22$
   from the SDSS DR8 on the sky in different coordinate systems. The
   top panel shows the map in right ascension and declination, the
   middle panel in Galactic coordinates, while the bottom in a
   coordinate system ($\Lambda$,$B$) aligned with the orbit of
   Sagittarius, as defined in \citet{Ma03}. Several stellar
   streams are clearly visible, the most prominent of which is the one
   originating from the Sgr dSph. The Sgr stream dominating the area
   around North Galactic Cap has been seen in the previous SDSS data
   releases. While some pieces of the southern stream have
   been revealed before, the new data gives a much more complete picture.
Similarly to the tail in the North, the tail in the
   South appears to have a fainter extension at one side (at higher B). The
   present location of Sgr dwarf is marked by a red star. The dashed
   red line is the projection of the Sgr orbital plane, as defined in
   \citet{Ma03}, and the blue dotted line shows the outline of the
   comparison field as discussed in the text.}
\label{fig:fos_dr8}
\end{figure*}

\section{Introduction}

The Milky Way has clearly not finished assembling, as the two
Magellanic Clouds are coalescing into the Galaxy. As stochastic
satellite infall continues in the Galactic halo, it gives us a chance
to bootstrap our cosmological theories of structure formation to the
local observables.  By measuring the signatures of accretion of
galactic fragments -- manifested in streams of stripped gas and stars
-- onto the Milky Way, we can study the Galaxy's underlying matter
distribution.

Of the Galactic satellites surviving to date, the Sagittarius (Sgr)
dwarf galaxy is one of the most massive, third after the LMC and SMC
(e.g. \citealt{Ni10}). It is however, not going to survive for much
longer. After its discovery by \citet{Ib94}, it was 
quickly realized that the Sgr dwarf was losing its stars to the Galactic
tides at a high rate (e.g. \citealt{Jo95, Ly95, Ma96}). It was
only when most of the sky was imaged by the 2MASS and the SDSS surveys
that the amount of damage done to Sgr became really apparent
(e.g. \citealt{Ne02,Ma03}). Currently, it is established that the
stellar debris torn from the dwarf wraps around the Galaxy at least
once, i.e. leading and trailing tails can be each seen covering over
$\pi$ radians on the sky \citep{Ma03,Be06}. Hundreds of stars in the Sgr tails
have
their radial velocities measured and some of these also have reliable
chemical abundances \citep{Ma04,Ch07,Mo07,Ya09,Ch10}. A number of star
clusters are believed to have originated in Sgr and are now left
free-floating in the Milky Way halo after having been torn from 
the disrupting galaxy (e.g. \citealt{Law10b}).

Unfortunately, the unbound stars (and star clusters) are not simply
tagged according to their past Sgr membership. Instead, objects are
typically classified as such based on their proximity to the Sgr
orbital plane, their heliocentric distances and radial velocities. Models have
had some considerable success explaining this data \citep[e.g.][]{LM10a}, but
difficulties remain. In particular, the
Virgo Over-density \citep{Ju08} and
the ``bifurcation'' of the leading tail \citep{Be06} are both examples
of substructures that lie close to the Sgr plane whose origins remain
obscure. Although there were attempts to include these in the picture of Sgr
disruption (e.g \citealt{Fe06, Ma07, Pe10}), it now seems that little 
progress has been made.

Finally, another stream (the Cetus stream) on a polar
orbit has recently been announced to overlap with Sgr debris in the
Southern Galactic hemisphere by \citet{Ne09}. Although this has a different
kinematical signature to Sgr, it coincides in space and metallicity. Untangling
the debris in the South is crucial to an understanding of the Cetus stream, as
well as the Sgr.

This is the first of two observational papers in which we report new
insights into the formation of the Sgr stream and its neighbouring
stellar halo substructures.  In a companion paper, we use the
multi-epoch observations of Sgr stars in Stripe 82 to measure the
proper motion of the stream~\citep[cf.][]{Ca12}.  Here, we revisit the
photometric data previously available from the Sloan Digital Sky
Survey (SDSS) \citep{Fu96,Gu98,Gu06,Yo00} archives as well as new 
measurements made public as part of the new Data Release 8
(DR8)\citep{Ai11,Ei11}. Crucially, this dataset now includes
significant coverage of the southern Galactic hemisphere not available
to \citet{Be06}.

The paper is arranged as follows. We extend the `Field of Streams'
plot~\citep{Be06} to the south in Section 2. This shows immediately
that the Sgr stream -- in the somewhat misleading nomenclature of our
earlier paper -- is bifurcated. Everywhere we look, in both the south
and the north, there is evidence for what appears to be two streams. In
Section 3, we use starcounts and Hess diagrams \citep{He24} to characterise the
density profiles and stellar populations of the streams. Where the
streams cross Stripe 82, we can take advantage of the coadded
photometry~\citep{An11}, which reaches $\sim$ 2 magnitudes deeper than
the single epoch SDSS measurements. We use photometric metallicities
to demonstrate that the two streams have different chemical
properties. Untangling the substructure is considerably complicated by
the existence of a further stream, already noticed by \citet{Ne09} and
dubbed the Cetus stream. This is studied in Section 4 using blue
straggler (BS) and blue horizontal branch (BHBs) stars.

\begin{figure*}
\begin{center}
\includegraphics[width=0.49\textwidth]{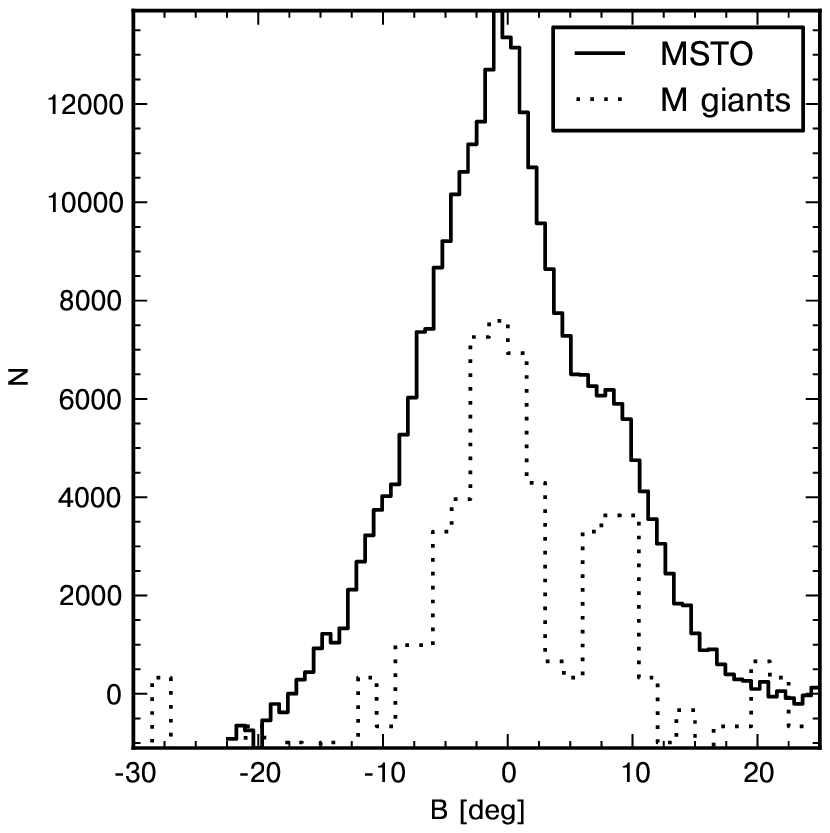}
\includegraphics[width=0.49\textwidth]{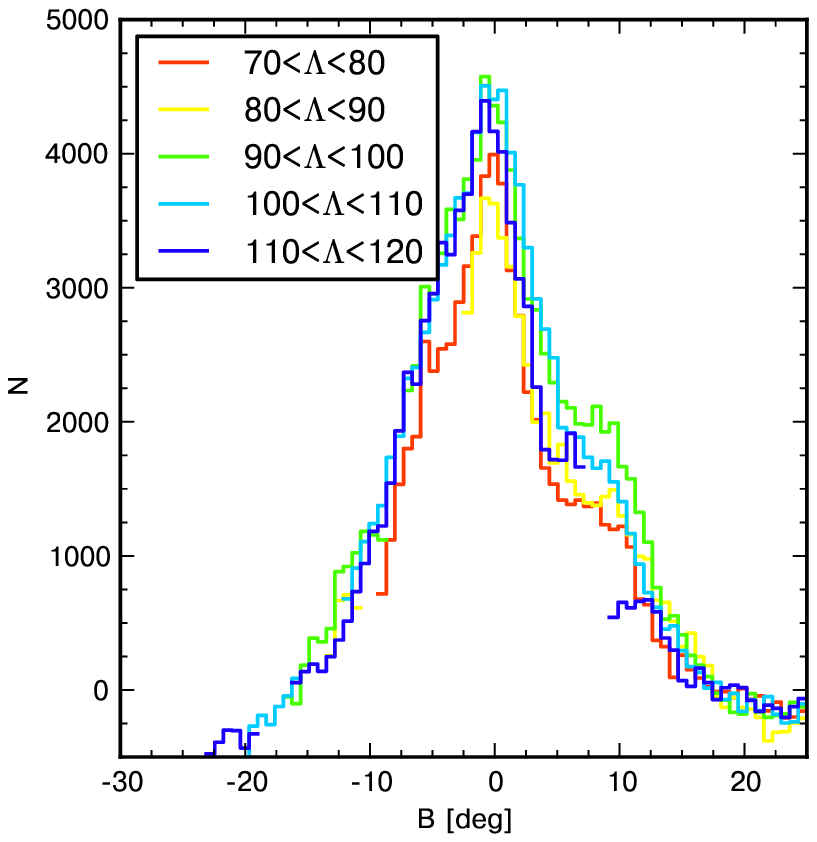}
\caption{Left: Density of MSTO stars (with the same
  color-magnitude selection as for Figure 1) across the Sgr stream in
  the south ($90^\circ <\Lambda< 120^\circ$) (solid line) and the
  density of 2MASS M giants in the same region (dotted line). A
  constant background (around 7000 stars per bin) has been subtracted from the
histograms.
Right  panel: The density of MSTO stars in different slices across the
  stream from $70^\circ <\Lambda< 80^\circ$ to $110^\circ < \Lambda<
  120^\circ$. Although the secondary stream seems to have the same
  offset from the main stream at different $\Lambda$, this is not
  actually the case as shown in Figure~\ref{fig:twoGauss}.}
\label{fig:crosssects}
\end{center}
\end{figure*}
\begin{figure}
\begin{center}
\includegraphics{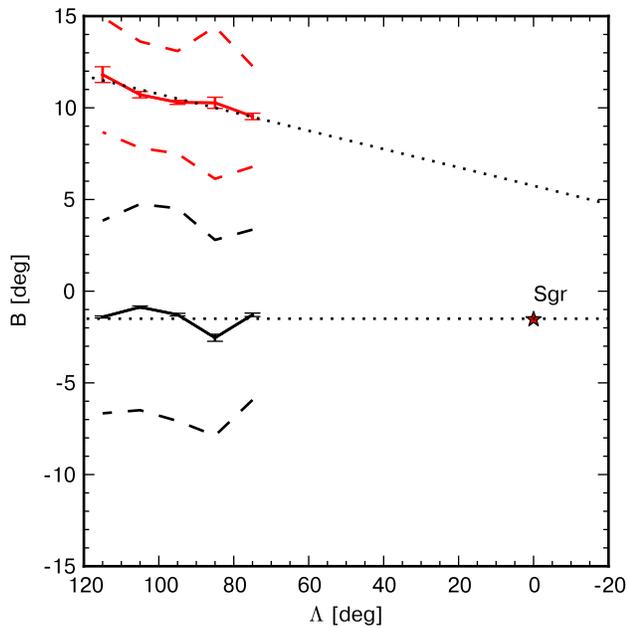}
\caption{Full lines show the centers of the bright and faint streams as a
  function
  of longitude $\Lambda$ as determined by a double Gaussian fit to the
  profile of Figure~\ref{fig:crosssects}. The dashed lines show the
  1 $\sigma$ widths of the streams, whilst the dotted lines show the
  extrapolation of the streams' centroids on approach to the Sgr
  remnant, which is marked by the red star.}
\label{fig:twoGauss}
\end{center}
\end{figure}

\begin{deluxetable}{ccccc}
\tabletypesize{\footnotesize}
\tablewidth{0pt}
\tablecaption{Locations and widths of the streams\label{tab:locs}}
\tablehead{ \colhead{$\Lambda$} &\colhead{$B_{cen,bright}$} 
&\colhead{$B_{cen,faint}$}  &\colhead{W$_{bright}$} 
&\colhead{W$_{faint}$}
\\
\colhead{deg}  &\colhead{deg}  &\colhead{deg}  &\colhead{deg}  &\colhead{deg}
}

\startdata
75&  -1.3&  9.5&  4.6&  2.7\\
85&  -2.5&  10.3&  5.3&  4.1\\
95&  -1.3&  10.3&  5.8&  2.8\\
105&  -0.9&  10.7&  5.6&  2.9\\
115&  -1.4&  11.8&  5.3&  3.1
\enddata

\end{deluxetable}

\begin{deluxetable}{ccccc}
\tabletypesize{\footnotesize}
\tablewidth{0pt}
\tablecaption{Distances to the bright stream\label{tab:sgr_dist}}
\tablehead{\colhead{$\Lambda$} &\colhead{$m-M$}
&\colhead{$\sigma(m-M)$} \\
\colhead{deg}  &\colhead{mag}  &\colhead{mag}  } 

\startdata

92.5& 16.69 &0.06\\
97.5& 16.72 &0.06\\
102.5& 16.86 &0.04\\
107.5 &17.01 &0.07\\
112.5 &17.17 &0.09\\
117.5 &17.31 &0.06\\
122.5 &17.27 &0.05\\
127.5 &17.28 &0.10
\enddata

\end{deluxetable}

\section{The Stellar Halo in the South}

To study substructure in the Galactic stellar halo, we select old and
moderately metal-poor stars with the simple color and magnitude cuts
$0 < {g}-{i} < 0.7$ and $19.5 < {i} <
22$. According to model isochrones, (e.g., \citealt{Gi04}), our sample
is dominated by the Main Sequence turn-off (MSTO) stars with
metallicity $Z \lesssim 0.02$ and absolute magnitude $3 \lesssim M_i
\lesssim 6$, occupying the range of heliocentric distances $10
\lesssim {\rm D (kpc)} \lesssim 60$.

The density of $\sim$ 13,000,000 stars that passed the above color and
magnitude cuts in the SDSS DR8 dataset is shown in
Figure~\ref{fig:fos_dr8} in equatorial and Galactic coordinates as well as in 
the coordinate system approximately aligned with Sgr orbit. The arc
of the Sgr tail -- note the two distinct streams, or branches A and B
in the notation of \citet{Be06} -- is clearly visible in the area
around the North Galactic Cap (NGC), as has been seen in the previous
SDSS data releases.  Also visible in the north are the Orphan Stream and
the Monoceros structure crossing the branches of the Sgr. DR8 reveals a large
continuous
portion of the Sgr tail in the Southern Galactic
hemisphere. Curiously, this tail too is seemingly accompanied by
another fainter stream following it at slightly higher declination. In fact this
is not the first sighting of this structure. \citet{Wa09}
showed that the density slice through the Sgr stream in the southern
Stripe 82 contains at least two maxima.

\begin{figure}
\begin{center}
 \includegraphics{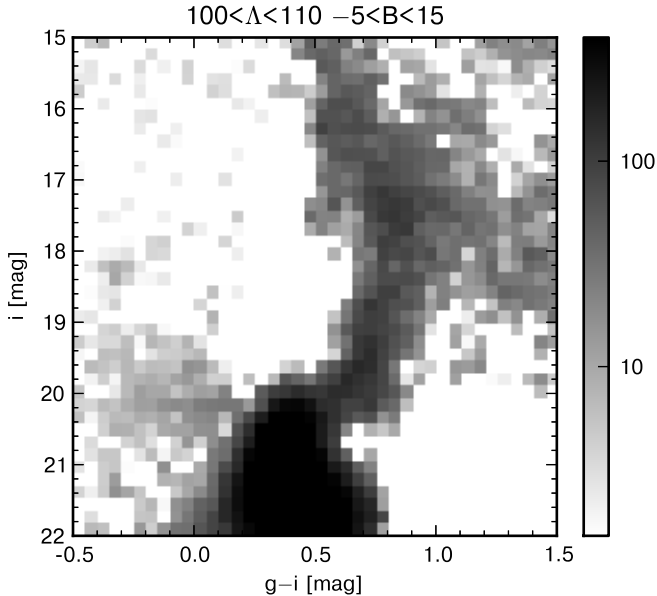}
 \caption{Background subtracted Hess diagram of the Sgr stream in
   the area defined by $100^\circ<\Lambda<110^\circ$ and
   $-5^\circ<B<15^\circ$. The background has been obtained from the
   symmetric area relative to the Galactic plane (which is marked by
   the blue dotted lines in Figure~\ref{fig:fos_dr8}). Multiple
   stellar evolutionary phases are clearly seen: MSTO, subgiants, red
   giant branch, blue stragglers and BHBs at $i \sim 18.2$. The curvy
   black region at bright magnitudes ($i\lesssim 19$, $g-i\gtrsim0.5$) is caused
by the imperfections of
   the background subtraction. On top of that problematic region, the
   red clump is located ($i \sim 17.5$).}
\label{fig:hess}
\end{center}
\end{figure}

\begin{figure}
\begin{center}
\includegraphics[width=3.39in]{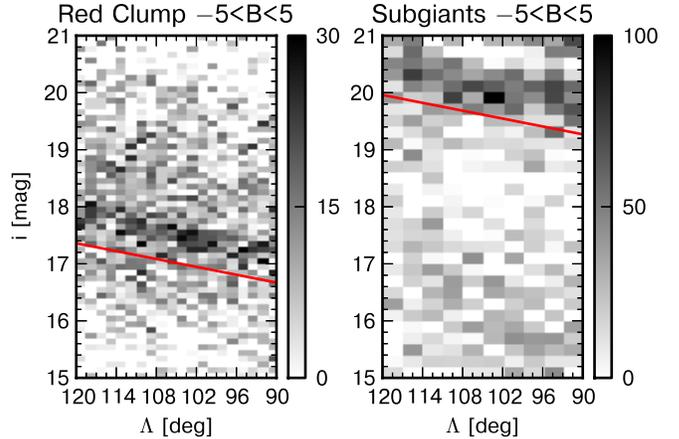}
\caption{
Measurement of distances and distance gradients along the stream using two
different tracers: sub-giant branch stars and red clump stars. Left:
2D-histogram of red clump star counts as a function of longitude along the
stream $\Lambda$ and $i$-band magnitude. The distance gradient is clearly visible.
Right: similar 2D-histogram for sub-giant stars, selected using a
combination of $g\!-\!i$ and $g\!-\!r$ colors.
The exact same gradient as for the red
clump is clearly visible. The red line on both panels shows the distance
gradient of 0.023 mag\,deg$^{-1}$ (offset for clarity).
}
\label{fig:distgrad}
\end{center}
\end{figure}

It is useful to define a heliocentric coordinate system aligned with
the Sgr stream. Such coordinate systems, whose equator aligns with the
stream, have already proved useful in similar studies \citep[e.g.][]{Ma03,Ko10}.
Using the notation of \citet{Ma03}, we introduce coordinates
($\Lambda, B$) given by their eqn (9). The equator of this spherical
cooordinate system coincides with the Sgr debris midplane. The bottom
panel of Figure~\ref{fig:fos_dr8} shows the data in this coordinate
system, with the Sgr debris now straddling the equator. This reveals
that, in the new projection, rather than a ``bifurcation'' of the Sgr
stream \citep[the somewhat misleading term introduced in][]{Be06}, the
stellar density appears resolved into two, sometimes superposed,
independent streams with seemingly different angular widths and
density profiles.

We are led to the conclusion that everywhere where the Sgr tidal
debris can be detected in SDSS there exists at least one additional density
component following or over-lapping the Sgr stream. In what follows,
we will attempt to empirically describe and untangle these structures by
examining their density profiles, distances and chemical abundances.

\begin{figure*}
\begin{center}
\includegraphics{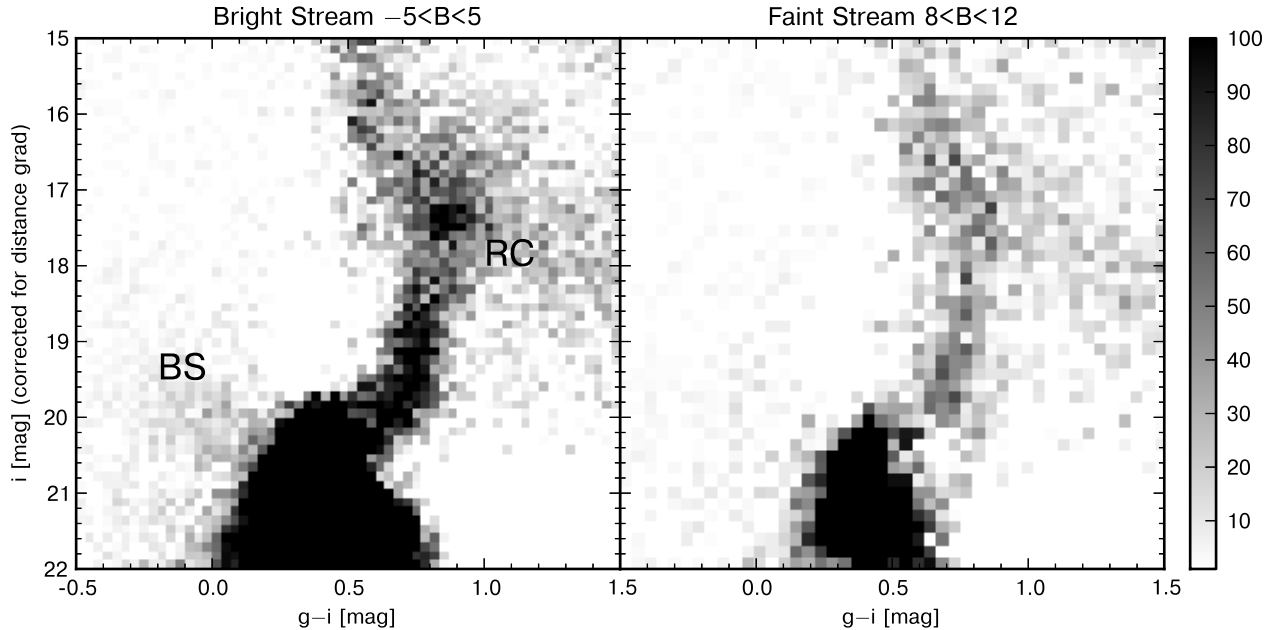}
\caption{Hess diagrams determined separately for the main Southern
  ($-5^\circ<B<5^\circ$) and the secondary Southern streams
($8^\circ<B<12^\circ$). In both
  cases, the Hess diagrams are corrected for the distance gradient of
  0.023 mag deg$^{-1}$. We have marked the red clump (RC) and blue
  straggler (BS) populations apparent in the brighter of the streams.}
\label{fig:superHess}
\end{center}
\end{figure*}

\section{The Sgr Stream in the South}

\subsection{Starcounts}

We begin by quantifying the difference in the centroids, widths and
density profiles of the streams visible in the South Galactic Region.  The left
panel
of Figure~\ref{fig:crosssects} shows the density of MSTO stars across
the stream in the region ($90^\circ<\Lambda<120^\circ$). For
comparison, we also show the density of M giants extracted from the 2
Micron All-Sky Survey (2MASS, see e.g., \citealt{Ma03}) using the 
cuts from \citet{Ma03}, namely $J-K_s>0.85$,
$0.22<J-H-0.561\,(J-K_s)<0.36$,
$10<K_s<12$. Both profiles show clear evidence for bimodality,
though it is unclear whether the two structures are distinct or overlapping. The
right panel of Figure~\ref{fig:crosssects} shows
cross-sections across the stream in different slices. As we march
along the stream, at least in the region $90^\circ <\Lambda<
120^\circ$, the cross-section remains quite invariant, but the
offset of the secondary stream from the main stream changes gently
with longitude.

Let us assume that the one-dimensional profile of each stream is a
Gaussian whose centroid and full width at half maximum may vary with
longitude. This simple model of the stellar density in a tidal
stream is of course not completely physical, but nevertheless should
be sufficient to describe the pieces of the streams in the south.  We
now extract the centroids and widths by fitting two Gaussians to
the starcount data. Figure~\ref{fig:twoGauss} shows their behavior as a
function of longitude $\Lambda$ along the stream. Perhaps
unsurprisingly, the tracks of the centers of two streams are not
exactly parallel, but they are slightly converging. Nonetheless, it
is surprising that the convergence point is not close to the Sgr
remnant (marked as a red star).  This gap decrease along the stream is
already a broad hint that the the two structures seen in
Figure~\ref{fig:crosssects} are separate streams following similar but
slightly different orbital paths, as opposed to a single stream with
substructure. In Table~\ref{tab:locs} we provide the data on the centroids of
the streams used for Figure\ref{fig:crosssects}.

Although \citet{Be06} used the term ``bifurcation'' to describe the Sgr
stream in the north, the new data and analysis suggests that the term
is misleading -- rather, in both the north and the south, the stellar
density appears to be resolved into two independent streams with
different angular widths.

Of course, we have made the assumption that the profiles are Gaussian. 
There is evidence from simulations that stream profiles can be lop-sided
\citep{LM10a}. With the current data, it is not possible to distinguish 
with absolute certainty between the hypotheses of two streams and a single
lopsided stream. Nonetheless, in our opinion, the presence of two peaks 
in the M giants, the detailed shape of the MSTO density profile and 
the behavior of centroids argues in favour of the hypothesis of two streams.

\begin{figure*}
\begin{center}
\includegraphics[width=.49\textwidth]{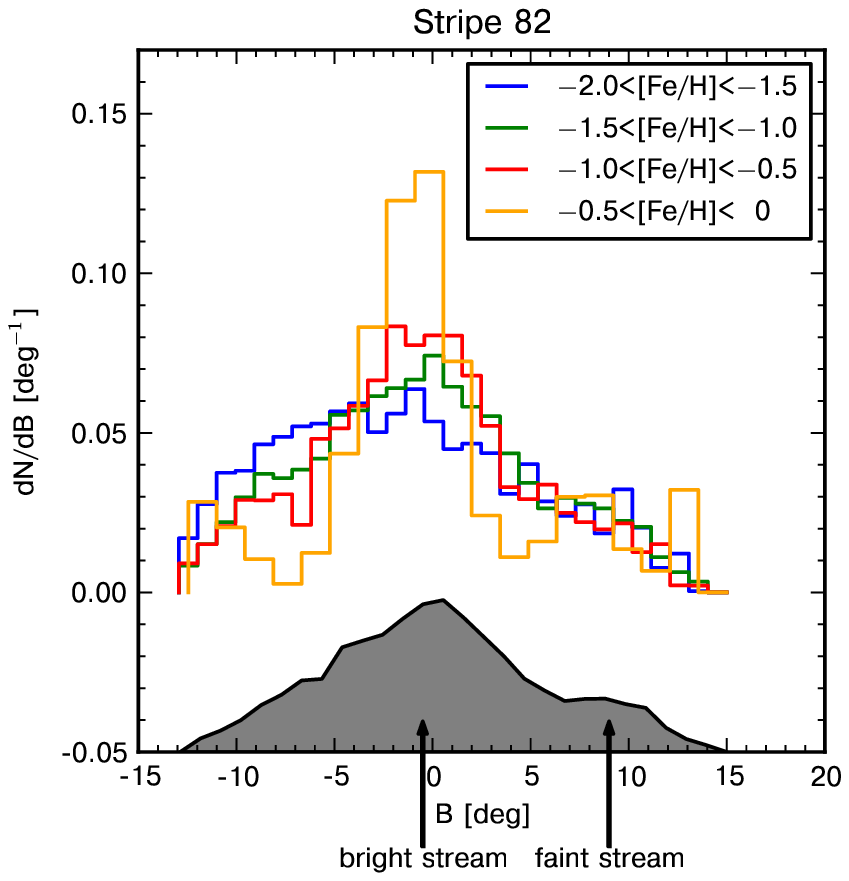}
\includegraphics[width=.49\textwidth]{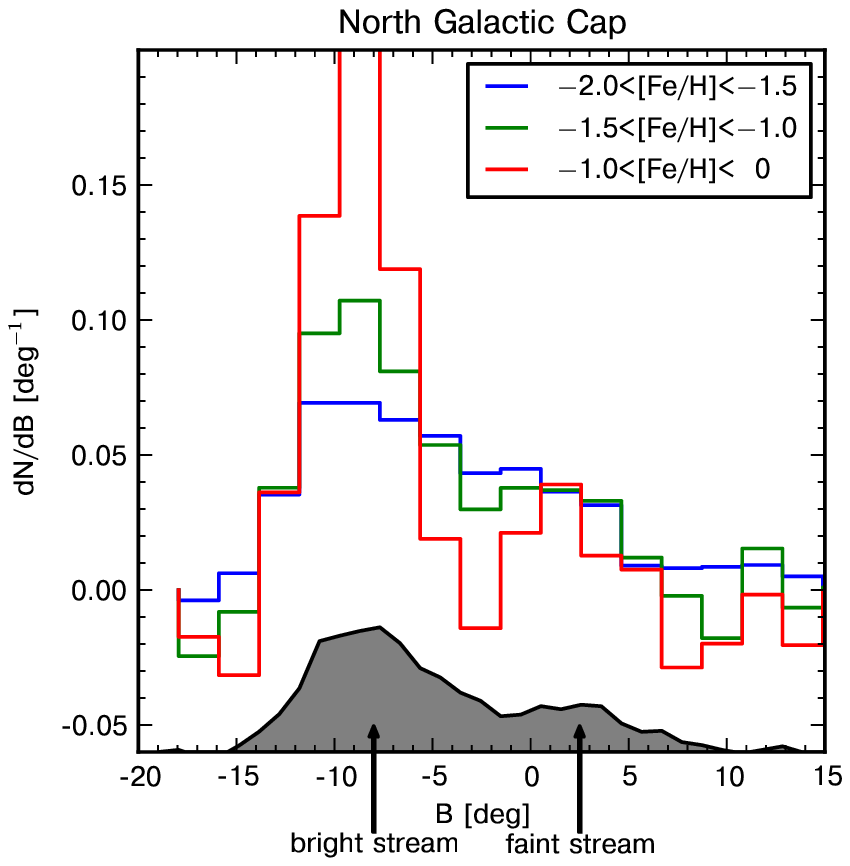}
\caption{Background-subtracted distributions of stars of different
  metallicities across the streams. In the left panel, the stars have
  been selected from the Stripe 82 coadded dataset, and in the right
  panel from the `Field of Streams' ($205^\circ<\Lambda<240^\circ$). The shaded
histograms show the density
  distribution of all MS stars without splitting in metallicity bins.
  Note that the two streams have
  different metallicity distribution functions, with the brighter stream
  containing substantial numbers of metal-rich stars as compared to
  the fainter stream.  This is true both for the streams in the
  south, and those in the north. The metallicities have been measured
  using the formula from \citet{Bo10}.}
\label{fig:metallicities}
\end{center}
\end{figure*}

\subsection{Color Magnitude Diagrams and Distance Gradients}

Distances to many different parts of the Sgr stream have been measured
in the past using various stellar tracers: Carbon stars
(e.g. \citealt{To98}), BHBs (e.g. \citealt{Ya00,Ne03}), sub-giant
branch stars (e.g. \citealt{Be06}), red-clump stars
(e.g. \citealt{Co10}) and RR Lyrae variables (e.g. \citealt{Pri09,
  Wa09}). However, when combined to provide as continuous a coverage of
the stream as possible, the results of these methods do not always appear to be
fully consistent. Distances to the stream in the south still rely on the
comprehensive study of M giants extracted from the 2MASS dataset
(e.g. \citealt{Ma03}).

Here, we will rely on the SDSS photometric data and concentrate on the
area in the Southern Galactic hemisphere where the stream is imaged
contiguously. Our aim is to construct clean Hess diagrams of the two
streams so as to analyse their stellar populations. Distances, or more
accurately, relative distances along the stream are needed. If
uncorrected for distance gradients, the features in our Hess diagrams
lose sharpness. Here, we will use red clump and subgiant
stars as distance indicators.

To construct the Hess diagrams, we make use of the fact that the
Galaxy is, to a good approximation, symmetric about the Galactic
plane. The blue dotted line in Figure~\ref{fig:fos_dr8} outlines
the area in the North corresponding to the main patch of Southern SDSS 
data mirrored in the Galactic plane. In Figure~\ref{fig:hess}, we 
show the Hess
diagram of the Sgr
streams in the range $100^\circ < \Lambda <110^\circ$ and
$-5^\circ<B<15^\circ$. We have subtracted the equivalent mirrored patch
as a proxy for the background region. The
existence of multiple stellar populations is immediately apparent from
the richness and thickness of the features in the Hess diagram.  We
can identify a fattened MSTO, subgiant and red giant branches, as well
as populations of BHBs and BSs.  Nonetheless, our background
subtraction is not perfect and is the cause of some graininess in
the figure, especially at brighter magnitudes. This is particularly
troublesome in the region of the red clump stars.

Some of the blurring and thickening of features is of course due to
the fact that the heliocentric distance is changing along the
streams.  Our next step is to measure the distance gradient, which we
quantify by studying two different tracer populations in
Figure~\ref{fig:hess}. We select red clump stars by the colour cut (cf.
\citealt{Co10})
\begin{equation}
  0.8 < g\!-\!i < 0.95.
\end{equation}
With a little more effort, we can select subgiant stars using the a
linear combination of $g\!-\!i$ and $g\!-\!r$ colors, namely
\begin{equation}
0.45<0.628\,(g\!-\!i)+0.529\,(g\!-\!r)-0.028<0.55.
\end{equation}
For both populations, we show 2D histograms of $i$ band magnitude
versus longitude along the stream $\Lambda$ in
Figure~\ref{fig:distgrad}.  We see that the gradient is linear to an
excellent approximation, and reassuringly the same for both
populations. In this area of sky, the longitudinal gradient is
$\sim$0.023 mag deg$^{-1}$. The distances measured along the brighter stream
from the the red clump population are shown in the Table~\ref{tab:sgr_dist}
under the assumption that $M_i=0.6$ \citep{Bela06}.

Having identified the relative distances of populations along the
stream, we can correct for the gradient and obtain cleaner Hess diagrams, as
shown in Figure~\ref{fig:superHess}. Here, the left panel refers to the brighter
stream $(-5^\circ < B< 5^\circ$), while the right panel to the
fainter stream $(8^\circ < B <12^\circ)$. It is noticeable that the
Hess diagram of the fainter stream shows much thinner sub-giant
and red-branch regions. Furthermore, it does not possess multiple
turn-offs and a prominent red-clump like the brighter southern
stream. To check that the latter is not an artefact caused by low number 
statistics in the faint stream, we modelled the ratio of red clump 
to MSTO stars using a decomposition of the density profile into two
Gaussians (as used for Fig.~\ref{fig:crosssects}).
With 95 \% confidence, this ratio
in the fainter stream is smaller than that in the brighter.
This suggests the existence of a simpler and more metal-poor
population in the fainter stream, and more complex and more
metal-rich population in the brighter stream.

We can confirm this result by making use of the Stripe 82 data. Stripe
82 has multi-band and multi-epoch imaging, which \citet{An11}
exploited to build a catalogue that reaches $\sim$ 2 magnitudes deeper
than the single epoch SDSS measurements. We compute photometric
metallicities using the formula provided by \citet{Bo10} and report
the results in Figure~\ref{fig:metallicities} both for Stripe 82 (left
panel) with $19 < r < 21.5$ and for the Sgr streams in the north between 
$205^\circ < \Lambda < 240^\circ$ with $19 < r < 21$.  Since Stripe 82 crosses
the Sgr streams at a significant angle, it is important to understand that
the left part of the plot corresponds to the $\Lambda\sim 110^\circ$
region while the right corresponds to $\Lambda \sim 50^\circ$, which
is much closer to the Sgr progenitor. Despite this limitation,
Figure~\ref{fig:metallicities} shows clearly that the brighter stream has
significant numbers of high metallicity stars, while the secondary stream has
fewer. 

\begin{figure}
\begin{center}
 \includegraphics{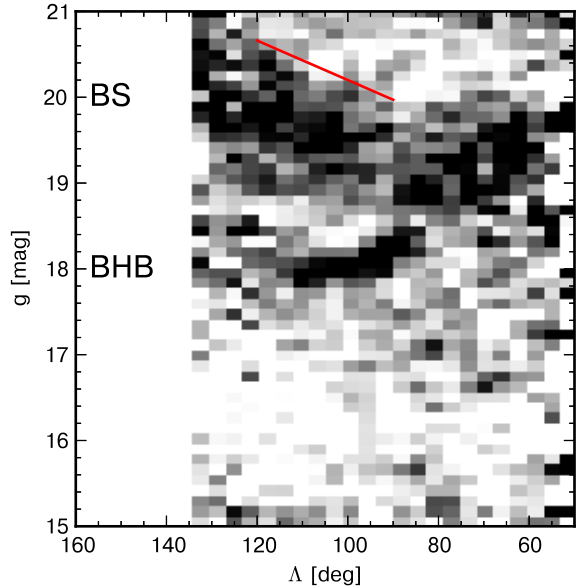}
 \caption{Magnitude distribution of BHB/BS-like stars as a function of
   the angle along the stream. The diagram shows two classes of
   objects coming from two structures: BS from Sgr stream with the
   same distance gradient as the subgiants/red clump (shown as an offset red
line), and BHBs from
   Cetus stream with an opposite distance gradient at $g\approx 18$ and
   $80^\circ\lesssim\Lambda\lesssim 120^\circ$.}
\label{fig:bhbs}
\end{center}
\end{figure}
\begin{figure*}
\begin{center}
\includegraphics[width=0.35\textwidth]{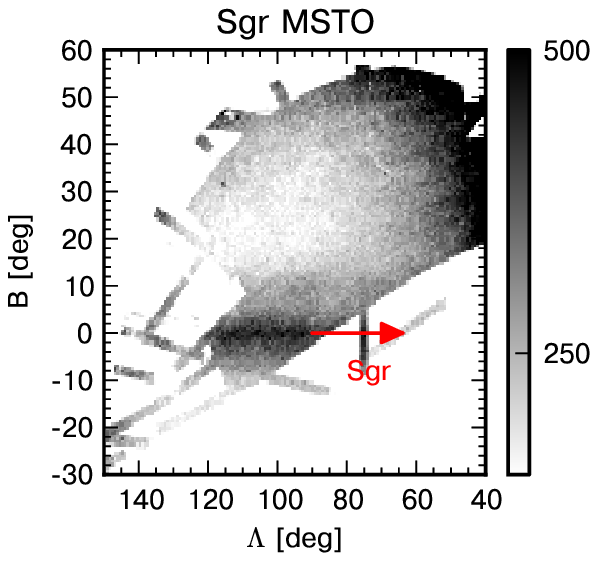}
\includegraphics[width=0.306\textwidth]{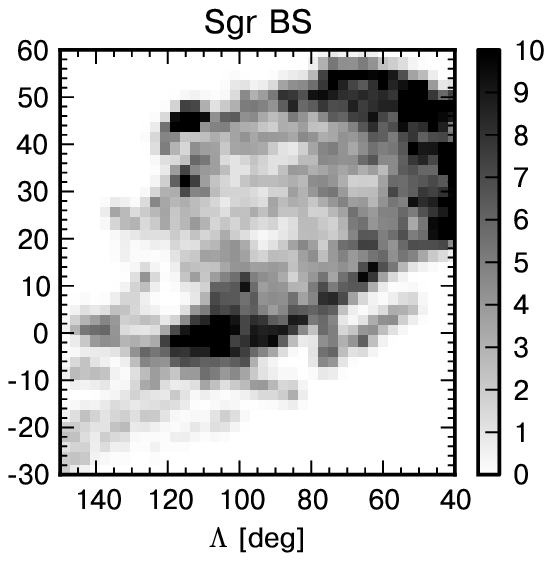}
\includegraphics[width=0.321\textwidth]{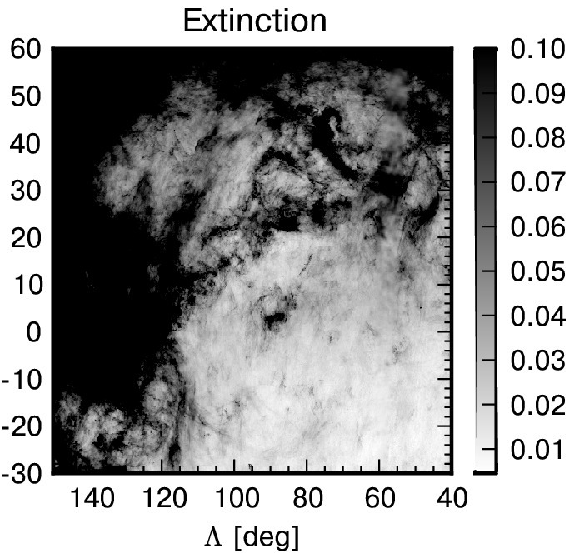}
\caption{
   Left: Density of high Galactic latitude ($|b| > 20^\circ$) MSTO stars
  selected using the same color-magnitude cuts as for Figure~\ref{fig:fos_dr8}.
The  arrow shows the direction of motion of Sgr stars.
  Middle: Density of high Galactic latitude BS stars
   $-0.3<g\!-\!r<0$, $0.9<u<1.3$ and  $18.5 <g < 20$ on the sky confirming
their
  attribution to the Sgr
  streams. The 1-D profile of the BS stars across the stream confirms
  that the BS stars are present in the secondary stream too (not
  shown). The slight dip in the density map visible at
  $\Lambda\sim85^\circ$ $B\sim5^\circ$ is caused by problems with the
  extinction correction. The peak at $\Lambda \sim 110^\circ$ $B\sim30^\circ$ is
  associated with M33. Milky way disk stars are also visible at lower latitudes.
Right: \citet{Sc98} extinction map for this region with the
elevated extinction (E(B-V)$\sim$ 0.1) patch at
$\Lambda\sim85^\circ$ $B\sim5^\circ$. 
}
\label{fig:bsstars}
\end{center}
\end{figure*}
\begin{figure*}
\begin{center}
\includegraphics[width=0.34\textwidth]{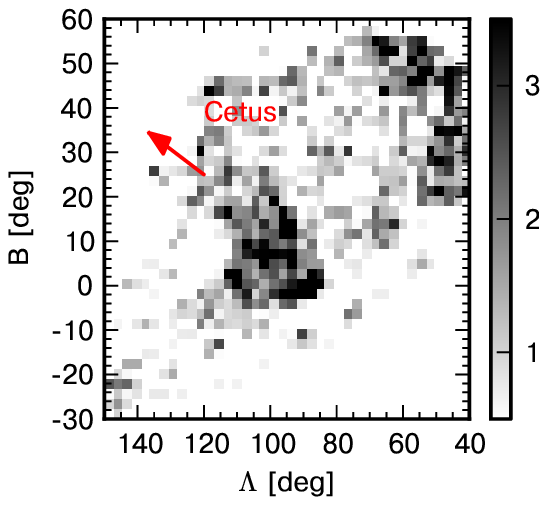}
\includegraphics[width=0.34\textwidth]{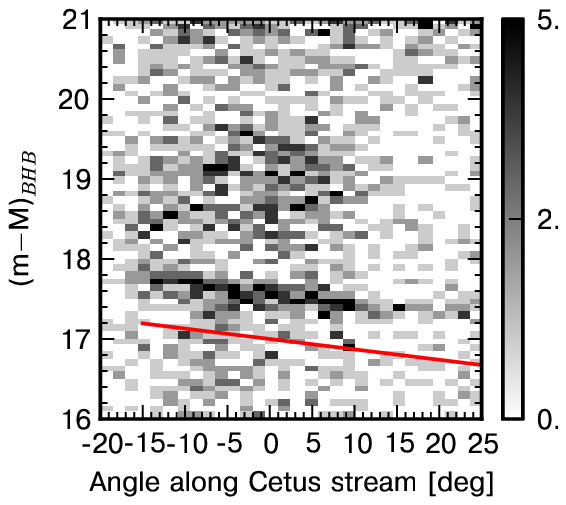}
\includegraphics[width=0.30\textwidth]{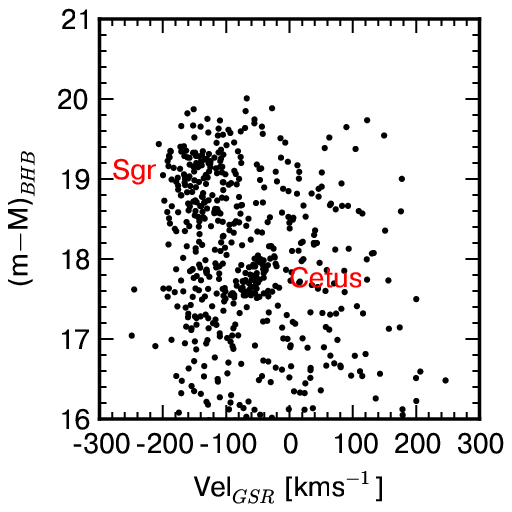}
\caption{
Left: Density of Cetus stream BHB stars on the sky. The stars were
  selected to be BHB-like according to standard $g\!-\!r$, $u\!-\!g$ cuts,
  Galactic latitude cut ($|b| > 20^\circ$),
  together with a position-dependant distance cut selecting predominantly Cetus
  stream stars. The arrow shows our measured distance gradient of Cetus stream
  BHBs. Middle: BHB stars in a coordinate system aligned with the
  Cetus stream, which is now clearly visible as an extended, narrow
  structure. The red line shows the distance gradient of
  0.013 mag\,deg$^{-1}$ offset for clarity. Right: The kinematical separation
  of the BS+BHB stars
  belonging to Sgr and Cetus stream for $80^\circ<\Lambda<120^\circ$ and
  $-10^\circ<B<20^\circ$. The velocity of Cetus stream stars in the Galactic
  rest frame is between $-$100 and $-$50 km\,s$^{-1}$, while Sgr stars have
  velocities between $-$200 and $-$100 km\,s$^{-1}$.}
\label{fig:bhbstwo}
\end{center}
\end{figure*}

\section{The Cetus and Sgr Streams}

\begin{deluxetable}{ccccc}
\tabletypesize{\footnotesize}
\tablewidth{0pt}
\tablecaption{Distances to the Cetus stream\label{tab:cetus_dist}}
\tablehead{ \colhead{$\phi_{1,Cetus}$} &\colhead{$m-M$} 
&\colhead{$\sigma(m-M)$} \\
\colhead{deg}  &\colhead{mag}  &\colhead{mag}  }
\startdata
-17.5 & 17.74 & 0.03\\
-12.5 & 17.73 & 0.01\\
-7.5 & 17.65 & 0.02\\
-2.5 & 17.54 & 0.02\\
2.5 & 17.51 & 0.02\\
7.5 & 17.43 & 0.01\\
12.5 & 17.35 & 0.02\\
17.5 & 17.34 & 0.02\\
22.5 & 17.39 & 0.04\\
\enddata

\end{deluxetable}

Whilst the picture so far is reasonably clear-cut, complications
emerge when we study the bluer populations, particularly the BHB and
BS stars. Of course, there is a long history of use of BHB stars for
studying structure in the stellar halo. The stars are relatively
abundant, and they occupy a narrow absolute magnitude range, which
makes them valuable distance indicators. However, there has been a
slight confusion as to the absolute magnitude of a typical BHB star in
the SDSS filter system: for example \citet{Ya00} and \citet{Ni10} have
used $M_g=0.7$, while \citet{Ne09} advocate the use of
$M_g=0.5$. It also known that the BHB absolute magnitude is also a function 
of color and metallicity. Fortunately, we do not need to address this 
issue here, as we will mostly use relative distances. 

Guided by the photometric properties of the globular cluster BHBs
studied by \citet{An08}, we choose to use the following simple color
cuts to select candidate BHB stars in the SDSS data: $0.9<u\!-\!g<1.3,
-0.35<g\!-\!r<0.0$.  At higher values of $g\!-\!r$, the contamination from
main sequence stars increases sharply. Of course, this color selection also
identifies out the BS stars \citep{Ya00}. We therefore might expect that
the
density distribution along the distance axis typically shows two
enhancements corresponding to BHB and BS stars separated by $\sim$ 2
magnitudes. Note that the BHBs have a narrow band of intrinsic
luminosities and so generate tight structures, whereas the BSs are 
poorer distance indicators and produce more diffuse structures~\citep[see
e.g., Figure 4 of][]{De11}.

Figure~\ref{fig:bhbs} shows the density of stars satisfying our colour
cut as a function of apparent magnitude and longitude. There are
indeed two evident structures present, but they do not correspond to
two roughly parallel density peaks which might be naturally
interpreted as BS and BHBs from the same structure. Rather the figure
forces upon us the interpretation that there are two classes of
objects arising from two distinct structures: the thick structure
possesses the same longitudinal distance gradient as the subgiant and
red clump stars, and this can be identified with BSs from the Sgr stream.
The other, thinner structure has the opposite distance gradient, and
we shall see that these are BHB stars lying in an entirely different
structure, namely the Cetus stream.

The discovery of the Cetus stream was announced by \citet{Ne09}. They
noticed a stream-like overdensity in low metallicity stars in SDSS DR7
that crosses the Sgr stream in the south at $b \sim -70^\circ$. The
Cetus stream can be distinguished from the Sgr on the basis of its
markedly lower ratio of BS to BHB stars and its different
kinematics. They also suggested that some BHB stars previously attributed
to the Sgr stream instead properly belong to the Cetus stream.

In the middle panel of Figure~\ref{fig:bsstars}, we display the density
distribution of stars on the sky satisfying the color cuts
$-0.3<g\!-\!r<0$ and $0.9<u\!-\!g<1.35$ together with $18.5 <g <
20$. The magnitude cut ensures that these stars are primarily BSs, and they
do indeed align with the Sgr stream along $B \approx 0^\circ$ (following the
MSTO distribution showed on the left panel of Figure~\ref{fig:bsstars}). There
is, however, some patchiness in the stellar distribution; this seems to be
associated with granularity in the extinction. We show the
extinction map of \citet{Sc98} in the right panel of
Figure~\ref{fig:bsstars}, and there is indeed an anomalous patch of high
extinction along the path of the stream at $\Lambda \approx 85^\circ,
B \approx 5^\circ$.

The spatial distribution of the Cetus stars in Figure~\ref{fig:bhbs} is
not easy to understand as the coordinate system is aligned with the
Sgr orbit. To reconstruct the orientation of Cetus, we first select
the BHBs in the range $17 < g< 18.5$ and $80^\circ< \Lambda < 130^\circ$
and $-5^\circ <B< 30^\circ$. We find the direction of the stream by
fitting a Gaussian to their magnitudes, allowing the center of the
Gaussian to vary linearly with $\Lambda$ and $B$. The fit yields
the magnitude of the BHBs as 
\begin{equation}
 g = -0.0112\,(\Lambda\!-\!100)-0.0064\,B\!+\!18.08. 
\end{equation}
together with the width of the Gaussian as $0.1$ mag. This allows us
to refine our selection of Cetus candidate members, as shown in the
left panel of Figure~\ref{fig:bhbstwo}. The arrow shows our measured
distance gradient of Cetus BHBs, which should lie roughly along the
extension of the stream. Despite low number statistics, the stream is clearly
visible and is at significant  angle to the Sgr stream. To confirm the
orientation of the stream, the middle panel shows the distance modulus of BHBs
as a function of angular distance along the Cetus stream. This angular
distance has been derived as longitude in the rotated coordinate system with
the pole at $(\alpha_p,\delta_p)=(294^\circ,30^\circ)$ and longitude
zero-point at $\alpha_0=25^\circ$. The absolute magnitudes of BHBs have 
been derived from g$-$r colors using Eq. 7 of \citet{De11}.
A narrow structure is visible spanning at least $40^\circ$. 
The measurement of these distances along the 
structure is given in Table~\ref{tab:cetus_dist}. 
We can further strengthen the case that
most of the BHB stars belong to the Cetus stream by studying their kinematics.
The right panel of Figure~\ref{fig:bhbstwo} shows the BS and BHB stars with
well measured velocities ($\sigma_V<$ 30 km\,s$^{-1}$) and  $-15^\circ<$ B
$<30^\circ$, $80^\circ <\Lambda<120^\circ$). The velocity of Cetus stars
corrected for Galactic rotation, $V_{\rm GSR}$ \footnote{This assumes a
Galactic rotation
velocity of 236 km\,s$^{-1}$ \citep{Bo09}, while the Sun's peculiar velocity
is taken from
\citet{Co11}} is between $-$100 and $-$50 km\,s$^{-1}$ while Sgr
stars have velocities between $-$200 and $-$100 km\,s$^{-1}$ (cf.,
\citealt{Ne09}). There is a clear and clean kinematical separation of the
BS and BHB stars belonging to Sgr and Cetus. Given the distance gradient and
radial velocity of Cetus, we see that its orbital motion is counter-rotating
with respect to that of Sgr.

\section{Conclusions}

We have studied the Sagittarius (Sgr) stream in the southern Galactic
hemisphere. In the SDSS Data Release 8, at most locations along the
orbit, an additional stream component can be discerned. As
evidenced from the density profiles of main sequence turn-off stars and M
giants, as well as the tracks of the components on the sky (see
Fig.~\ref{fig:crosssects}), we think that the most natural interpretation is
that there are two streams. Our modelling of the
cross-sections of the density suggests the existence of a
thicker brighter stream and a thinner fainter stream,
offset by $\sim 10^\circ$. The streams differ in integrated luminosity by a
factor of 5--10. There is also strong evidence that the two streams have
different metallicity distribution functions.

We used red clump and subgiant stars to measure the distance gradients
along the streams. This enables us to construct composite Hess
diagrams to study the stellar populations in the streams. The brighter
stream shows evidence for multiple turn-offs and a prominent red
clump, whereas the secondary stream does not. This suggests that the
brighter stream is composed of more than one stellar population, and
contains a significant number of metal-rich stars, much like the Sgr
remnant itself. By contrast, the fainter stream is dominated by a
metal-poor population. This conclusion is also supported by our
photometric metallicities computed for the region where the Sgr
streams cross the ultra-deep SDSS Stripe 82 coadded data.

Our analysis of the new data allows us to untangle a complicated mix
of tidal debris around the South Galactic Cap, where the Sgr streams
are crossed by the Cetus Stream at an angle of $\sim 30^\circ$. The
Sgr and Cetus streams have similar distances, though their stellar
populations and kinematics are different. The structures are not part
of the same disruption event, as the Cetus stream is counterrotating
with respect to the Sgr. On the basis of their density distribution
and kinematics, we have shown that most of the BSs belong to the Sgr
stream, whereas most of the BHBs belong to the Cetus Stream, in this
part of the sky. The BS to BHB ratio in the two streams is strikingly
different, as already pointed out by \citet{Ne09}.  Good
spectroscopic coverage of the area is essential to disentangle the
multiple overlapping streams with different chemical properties,
following different distance gradients.

The work in this paper has extended the `Field of
Streams'~\citep{Be06} to the south. The new imaging data show that,
just as in the north, the Sgr stream is accompanied by a fainter
stream. The right-hand panel in Fig~\ref{fig:metallicities} shows that
the population mix in the fainter stream of the ``bifurcation'' around the
North Galactic Cap does not contain as many metal-rich stars as the
brighter stream. The simplest explanation is that the southern faint
stream is part of the same structure as the northern faint stream.

These results raise the question: Is it possible to produce the streams with
the properties described in this paper in a disruption of one galaxy
or is more than one progenitor necessary?

Recently, two possible scenarios, both in context of the Sgr dSph
disruption, have shown how to form two almost parallel tidal streams
from the debris of one parent galaxy. Although \citet{Fe06} did not fully
match the available data, they did suggest that
branches A and B (see \citealt{Be06}) could be reproduced by multiple
wraps of the same stream offset on the plane of the sky by small
amount of differential precession. In this picture, the fainter stream
in the North Galactic Cap area corresponds to the dynamically old
tidal debris in the Sgr trailing arm. However, this mechanism does not
produce two distinct streams in the south and now seems to be ruled out. 

Alternatively, \citet{Pe10} point out that if the Sgr progenitor had a
rotating stellar disk misaligned with respect to its orbit, then
stripping naturally produces bifurcated debris tracks on the sky. However, the
model seems to be ruled out as it has been difficult to find strong evidence for
residual rotation in the remnant of the Sgr dwarf \citep{Pe11}. Moreover, in
the scenario of \citet{Pe10} the two streams are not expected to differ
significantly in their stellar populations content. This prediction is somewhat
difficult to reconcile with the new data on the stellar populations of the
streams. In some other models although, such as \citet{LM10a} the streams that
form at consecutive pericentric passages could have somewhat different
metallicity distributions, as metal-rich stars could have been torn
predominantly later, from deeper within the gravitational potential of the
satellite.

Finally, it is tempting to suggest that the two streams with
different properties have actually originated from two different
progenitors. The infall of satellites in groups is not particular but
general, as best illustrated by the recent arrival of the Large and
Small Magellanic Clouds. Cosmological simulations of structure
formation also find plenty of evidence for group infall
(e.g. \citealt{Li08}). The picture painted by the data looks more complex
than this and remains a challenge to understand. Although there has been
substantial progress in modelling the Sgr \citep[see e.g.,][]{Fe06, Pe10,
LM10a}, it remains true that there is no explanation of the nature of the two
Sgr streams.

\acknowledgements{SK acknowledges financial support from the Science
  and Technology Funding Council of the United Kingdom, whilst VB and MG
  thanks the Royal Society for the award of a University Research
  Fellowship. GFL thanks the Australian Research Council for support through
his Future Fellowship (FT100100268) and Discovery project (DP110100678).
  Most of the data processing
  has been done using the python programming language and the following open
  source modules: numpy\footnote{\url{http://numpy.scipy.org}},
  scipy\footnote{\url{http://www.scipy.org}}, matplotlib\footnote{\url{
  http://matplotlib.sf.net}}. The SDSS data was accessed using local
  SDSS copy stored in PostgreSQL database and Q3C module \citep{Ko06}.
Funding for SDSS-III has been provided by the Alfred P. Sloan Foundation,
the Participating Institutions, the National Science Foundation, and the
U.S. Department of Energy Office of Science. The SDSS-III web site is
http://www.sdss3.org/.

SDSS-III is managed by the Astrophysical Research Consortium for the
Participating Institutions of the SDSS-III Collaboration including the
University of Arizona, the Brazilian Participation Group, Brookhaven
National Laboratory, University of Cambridge, University of Florida, the
French Participation Group, the German Participation Group, the Instituto de
Astrofisica de Canarias, the Michigan State/Notre Dame/JINA Participation
Group, Johns Hopkins University, Lawrence Berkeley National Laboratory, Max
Planck Institute for Astrophysics, New Mexico State University, New York
University, Ohio State University, Pennsylvania State University, University
of Portsmouth, Princeton University, the Spanish Participation Group,
University of Tokyo, University of Utah, Vanderbilt University, University
of Virginia, University of Washington, and Yale University.}

\end{document}